\documentclass[11pt]{article}

 \def\beq{\begin{equation}}
 \def\eeq{\end{equation}}
 \def\lf{\left(}
 \def\rg{\right)}
 \def\lq{\left[}
 \def\rq{\right]}
 \def\lGr{\left\{}
 \def\rGr{\right\}}
 \def\p{{\partial}}
 \def\vb{{\vec b}}

 \def\vj{{\vec j}}
 \def\vk{{\vec k}}
 \def\vh{{\vec h}}
 
 \def\vx{{\vec x}}
 \def\vy{{\vec y}}
 \def\vA{{\vec A}}
 \def\vD{{\vec D}}
 \def\Db{{\bar D}}
 \def\vE{{\vec E}}
 \def\vG{{\vec G}}
 \def\vP{{\vec P}}

 \def\hp{{\hat{\psi}}}
 \def\hpd{{\hat{\psi}^\dagger}}
 \def\hc{{\hat{c}}}
 \def\hcd{{\hat{c}^\dagger}}
 \def\vj{{\vec{\jmath}}}
 \def\vnabla{{\vec\nabla}}

 \newcommand\half{{\scriptstyle{\frac{1}{2}}}}

\begin{document}

 \setlength{\baselineskip}{16pt}
 \title{Symmetries of Non-relativistic Field Theories\\ on the Non - Commutative 
 Plane\footnote{
 To appear in the Proceedings of the International Workshop
 {\it Nonlinear Physics} III. Talk presented by LM.
 To be published in {\it Theor. Math. Phys}.}}
 \author{
 P.~A.~Horv{\'a}thy $^{(a)}$, L. Martina $^{(b)}$, P.~C.~Stichel
 $^{(c)}$
 \\
 \small{ (a) {\it Lab. Math. Phys. Th\'eor.,  Universit\'e de
 Tours, P.  Grandmont, F-37 200 TOURS (France)}}
 \\
 \small{ (b) {\it Dip. Fisica and Sez. INFN, Universit\`a di Lecce,
 v. Arnesano, CP. 193, I-73 100 LECCE (Italy).}}
 \\
 \small{ (c) {\it An der Krebskuhle 21, D-33 619 BIELEFELD
 (Germany) }}}

\maketitle

 \begin{abstract}
     New developments on non-relativistic field theoretical models on
     the non commutative plane are reviewed.
      It is shown in particular that Galilean invariance strongly
      restricts the admissible interactions.
      Moreover, if a scalar field is coupled to a Chern - Simons 
      gauge field,
      a geometrical phase emerges  for vortex - like solutions,
      transformed by Galilei boosts.
\end{abstract}

\texttt{hep-th/0411139} 

 Non commutative solitons - finite action solutions
 of the classical equations of motion of noncommutative field
 theories - have attracted  great interest in the last few years,
 mainly in connection with strings and brane dynamics \cite{Szabo}.
 However, at very low energy (i.e. in condensed matter physics),
 the analysis of the Fractional Quantum Hall Effect (FQHE)
 \cite{Laughlin}, has suggested that the phenomenology can be
 expressed in terms of quasi-particles, related to states of
 strongly correlated electrons in the Lowest Landau Level. These
 quasi-particles are imbedded into an effective gauge connection
  of entirely quantum mechanical nature, related to
  Berry's phase \cite{Arovas}. As a consequence, the
 quasi - particles (anyons) have fractional statistics. In a
 field theoretical approach, the topological origin of this effect
 is encoded into  a non-dynamical vector potential $\vA$ which
 reproduces an Abelian Chern - Simons term in the action, minimally
 interacting with the massive field $\psi$ of a Landau - Ginzburg
 theory.  The original quasiparticles re-emerge as
 vortex like solutions, which carry fractional electric charge and
 unit magnetic flux.

 On the other hand, in the limit of vanishing mass,
 $m_e \rightarrow 0$, the classical Lagrangian
 for a system of interacting charged particles (electrons) in
 the plane becomes
 first order in time derivatives, providing us with
 hamiltonian equations
 of motion for non commuting variables \cite{Girvin}. Analogous
 Lagrangian (and also hamiltonian \cite{HMSanomal})  equations can
 be obtained for the mean values of the position and momentum for
 wave packets in magnetic Bloch bands \cite{Niu}. A Mead - Berry
 connection \cite{Bohm}, depending on the quasi - momentum,
 appears also in this case. It again encodes a geometric phase into the
 semiclassical description of the microscopic system.

 Similar, simplified particle models in the plane were
 introduced on a purely axiomatic mechanical setting in \cite{LSZ},
 resorting to an acceleration - dependent Lagrangian, and in
 \cite{DH}, where a formalism {\it {\`a} la
 Souriau} was used. More interestingly,
 fine-tuning  the magnetic field with respect to a new parameter $k$
 specific to the particle provides us with
 equations of motion, which reduce to the Hall equations for a
 charged particle in perpendicular electric and a magnetic fields.

 Three phenomena arise: i) the
 dimensional reduction of the phase space, ii) the non
 commutativity of the reduced "configurational" variables, iii) the
 parameter $k$ generates a second central extension of the
 Galilei group in the plane. This can be seen by computing the
 Poisson brackets of the boost generators
 \begin{equation}
      \left\{
      {\emph{G}}_{1},{\emph{G}}_{2}\right\}= k.
      \label{exoticcharge}
 \end{equation}
 Some time ago it has  been proved in fact \cite{exotic} that the Galilei
 group in 2+1 dimensions admits a 2-dimensional central extension,
 in contrast with the usual one, associated with the particle mass.
 However, its physical meaning remained obscure for a long time and
 the result was considered a mere mathematical curiosity.
  Thus, in order to reproduce the properties of the Laughlin
 quasiparticles, it is natural to
 consider a Chern - Simons field theory on a
 non commutative plane (NC-plane) as a better description of the
 FQHE. This idea was stressed in \cite{Susskind}. The
 NC-plane is represented as the $C^*$- algebra of the bounded
 operators generated by the Heisenberg algebra
 \begin{equation}
 \left[{\hat x}_i, {\hat x}_j\right] =  - \imath \varepsilon_{i
 j}\, \theta, \qquad \left(i,j = 1,2\right)
 \end{equation}
 where $\theta$ is a characteristic scalar parameter, playing the
 same role as $\hbar$ in the phase space, and $\hat{\varepsilon} =
 \lf \varepsilon_{i j}\rg $ represents the antisymmetric tensor in
 two dimensions.

 There exists a one-to-one mapping
 between the space $\mathcal{S}$ of the schwartzian functions
 $\psi$ on $\bf{R}^2$ and the $C^*$-algebra. It is  defined by the Weyl
 quantization formula 
 $$\hat{\psi}
 = \int \psi \left( {\vx} \right)\hat{\Delta}\left( \vx \right) d^2
 \left( \vx \right), 
 \quad\hbox{where}\hat{\Delta}\left( \vx \right)=
 \frac{1}{\left(2\pi\right)^2}\int e^{\left(
 \imath\vk\cdot\left(\hat{{ \vx}} - \vx\right)\right)} d^2 \vk$$
 is the point-like quantizer operator. The inverse is given by the
 Wigner de - quantization formula $\psi \left( \vx \right) =
 \textrm{Tr}\left( \hat{\psi} \hat{\Delta}\left( \vx \right)\right)
 $, where the translation invariant trace map $\textrm{Tr}\left(
 \hat{\psi} \right) = \int \psi \left( \vx \right)d^2  \vx $ can be
 introduced. It satisfies  the relation $\textrm{Tr}\left(
 \hat{\Delta}\left( \vx \right) \hat{\Delta}\left( \vy \right)
 \right) = \delta\left( \vx - \vy \right)$.  Thus, one is lead to a
 new associative non abelian algebra in the space $\emph{S}$  in
 terms of the Moyal ($\star$) product
 \begin{equation}
 \psi \star \varphi \left( \vx\right) = \textrm{Tr}\left(
 \hat{\psi} \hat{\phi} \hat{\Delta}\left( \vx
 \right)\right).
 \end{equation}
  This result allows us to rephrase any
 field theory, defined by an action
 \begin{equation}
 S\left[\hat{\psi}_\alpha\right] = \int dt \;Tr\left[ \rm{L} \lf
 \hat{\psi}_\alpha, \lq \hat{\p}_i , \hat{\psi}_\alpha \rq, \dots ,
 \rg \right]= \int dt d^2 \vx \mathcal{L}\lf {\psi}_\alpha, {\p}_i
 {\psi}_\alpha , \dots , \rg \label{genaction}
 \end{equation}
  for the operators in $C^*$, in terms of a nonlocal
 Lagrange density, involving the classical fields $\psi_\alpha$,
 their derivatives and their $\star$-products.

  The noncommutative
 version (see \cite{Susskind,LMS,BakCS}) of the non relativistic
 scalar field theory ($ m=1, \, e=1$ ) coupled to the Chern-Simons
 gauge field is given by the Lagrange density
 \begin{equation}
     \mathcal{L} =
      i\bar{\psi}\star D_{t}\psi-\half\overline{\vD\psi}\star\vD\psi
      +\kappa\left(\half\epsilon_{ij}
      \p_{t}A_{i}\star A_{j}+A_{0}\star F_{1 2}\right) -
      V\left( \psi, \bar{\psi}\right).
      \label{Clag}
   \end{equation}
   In (\ref{Clag}) one has introduced the $\star$ - covariant derivative and the
   $\star$ field - strength
 tensor
 \begin{eqnarray}
      D_{\mu}\psi=\p_{\mu}\psi-i  A_{\mu}\star\psi,\;\;
      \overline{D_{\mu}\psi}= \p_{\mu}\bar{\psi}+i \bar\psi\star(A_{\mu}) \label
 {lcovder}
      \\
      F_{\mu\nu}=\p_{\mu}A_{\nu}-\p_{\nu}A_{\mu}
      -i\big(A_{\mu}\star A_{\nu}-A_{\nu}\star A_{\mu}\big),
      \label{fieldstrength}
 \end{eqnarray}
  respectively. According to (\ref{lcovder}) the matter field
 $\psi$ is in the fundamental representation of the gauge group
 $U(1)_{*}$,  i.e.  the locally gauged fields are given by
 $$
 \tilde{\psi}=e^{\imath \lambda\left(\vx\right)}\star \psi,
 \quad
 \tilde{A_{\mu}}= e^{\imath \lambda\left(\vx\right)}\star
 \left(A_{\mu}+ \imath\partial_{\mu}\right)\star e^{-\imath
 \lambda\left(\vx\right)},
 \quad
 \tilde{F_{\mu \nu}}= e^{\imath
 \lambda\left(\vx\right)}\star F_{\mu \nu}\star e^{-\imath
 \lambda\left(\vx\right)}.$$
 A remarkable feature of the $U(1)_{*}$
 gauge theory (\ref{Clag}) is the quantization of the coupling
 constant \cite{kappaquant}
     $ \kappa=\frac{n}{2\pi},
      \,n \in \bf{Z} $, corresponding to the quantized filling factor
      in the FQHE \cite{Susskind,kappaquant}.
  Vortex - like solutions of such a model were discussed in
 \cite{LMS,BakCS}.

 As shown in (\ref{genaction}), the Lagrangian can be expressed as
 a trace over a Hilbert space, namely
 \begin{eqnarray}\label{NCLGOp}
 \rm{L} =  \imath \pi \kappa \textrm{Tr}\lq K^\dagger D_t K-K D_t
 K^\dagger\rq -2 \pi \kappa \textrm{Tr}\lq A_0\rq + \\ \nonumber 2
 \pi \theta \textrm{Tr}\lq \imath \hpd D_t\hp -\frac{1}{2
 \theta}\lf D\hp \lf D\hp\rg^\dagger + \Db \hp \lf \Db
 \hp\rg^\dagger \rg +  V\lf \hp ,\hpd \rg \rq ,
 \end{eqnarray}
 where we have redefined the gauge field operators as
 \begin{equation}\label{gK}
   K = \frac{1}{\sqrt{2 \theta}} \lf \hat{x}_1 - \imath \hat{x}_2 - \imath
 \theta
    \lf \hat{A}_1
 \imath \hat{A}_2 \rg \rg = \hat{c} - \imath \sqrt{2 \theta}
 \hat{A}_+ \label{K}
 \end{equation}
 and the corresponding adjoint $K^\dagger = \hcd + \imath \sqrt{2
 \theta} \hat{A}_-$. The operator $\hc =\frac{1}{\sqrt{2\theta}}\lf
 \hat{x}_1 - \imath \hat{x}_2 \rg $ and its adjoint $\hcd$ satisfy
 the canonical commutation relation $\lq \hc, \; \hcd \rq = 1$,
 furthermore in terms of complex variables $\lf z, \bar{z} \rg$ one
 has the representation $\lq \hc, \cdot \rq = \sqrt{2\theta}
 \p_{{z}}, \, \lq \hcd, \cdot \rq = - \sqrt{2\theta}\p_{\bar z} $ .
  The covariant derivatives act as
 \begin{equation}\label{Dt}
 D_t \hp  = \p_t \hp - \imath \hat{A}_0 \hp, \qquad  D_t \textsf{O}
 = \p_t \textsf{O} - \imath \lq \hat{A}_0, \textsf{O}\rq ,
 \end{equation}
 \begin{equation}\label{Dzf}
  D \hp     =  \sqrt{\frac{\theta}{2}} \lf D_1 - \imath D_2 \rg \hp
  =K \hp - \hp
 \hc
   ,
  \qquad \Db \hp  = \sqrt{\frac{\theta}{2}} \lf D_1 + \imath D_2 \rg \hp =
  \hp \hcd - K^\dagger \hp
 \end{equation}
 \begin{equation}
  D \textsf{O}  = \lq K, \textsf{O}
  \rq  , \qquad \Db \textsf{O}  =  \lq \textsf{O}, K^\dagger \rq ,
 \end{equation} for any vector operator $\textsf{O}$.

 Explicit,  and  at
 quantum level renormalizable, solutions can be found \cite{Baketal}
 for the (ungauged) fourth-order self-interacting
 Landau - Ginzburg model
  \begin{equation}
      \mathcal{L}={\emph L}_{0}-V^\star
      =\left(i\bar{\psi}\star \,\p_{t}\psi+
      \bar{\psi}\star \frac{\bigtriangleup\psi}{2}\right)
      -\frac{\lambda}{2}\,
      \bar{\psi}\star\bar{\psi}\star\psi\star\psi.
      \label{Baction}
 \end{equation}
 In particular, two-particle bound states were found.
 They are characterized by a dipolar length
 proportional to the transverse \emph{total momentum}, and the
 parameter $\theta$. The latter signals the breaking of galilean
 symmetry. This behaviour  is analogous to what found in certain
 string models.  The question of  galilean
 invariance in  NC-theories is addressed in
 \cite{HMS1,HMS2}. Below we review  and  amplify the
 results, reported in \cite{HS,Had}, on boost - invariant solutions
 in the NC-plane.

 In our attempt to perform a systematic symmetry analysis of  field
 theories on the NC-plane, we start with observing that for the free
 version of  (\ref{Baction}) $L_0$ is quasi - invariant and
 acquires divergence-like terms (possibly  proportional to
 $\theta$) with respect to  an ``exotic '' version of the
 10-dimensional Schr{\"o}dinger symmetry algebra. Note that because
 of the bilinear form of the Lagrangian and the integral property
 $$\displaystyle\int\! f\star g
 \left(\vx\right) d^2\vx =\int\!
 f\left(\vx\right) g\left(\vx\right)d^2\vx,
 $$
  the action concides with the usual one in the commutative plane.
 The Euclidean  subgroup is implemented by  the inner automorphisms
 \begin{eqnarray}\label{Eu} \psi_{trasl} = & \psi\left(\vx - \vh, t\right)= &
  e^{-\imath
 \frac{\hat{\varepsilon}\vh}{\theta}\cdot\vx} \star\psi\left(\vx,
 t\right)\star e^{\imath
 \frac{\hat{\varepsilon}\vh}{\theta}\cdot\vx},
 \\ \nonumber
 \psi_{rot} = & \psi \lf \mathcal{R}_\varphi^{-1} \vx , t\rg & ,
 \end{eqnarray}
 where $\vh$  and $\varphi$ represent  parameters of the
 translations $\vx \rightarrow \vx + \vh$ and of the rotations $\vx
 \rightarrow \mathcal{R}_\varphi \vx$, respectively. These relations
 express i) the nonlocality of the theory related to the scale of
 the momentum $\hat{\varepsilon}\vh$, ii) the emergence of the
 space translations as particular gauge transformations. This
 allowed us to express covariant derivatives in terms of the gauge
 field in  (\ref{Clag}) using "covariant fields"
 (\ref{K}) (see \cite{Szabo}).

  The usual one-parameter centrally extended
 Galilei transformation  is replaced by  ``exotic''
 two-parameter ones with (for $m = 1$) infinitesimal and finite expressions
 \begin{equation}
      \delta^{*}_{}\psi=(i\vb\cdot\vx)\star\psi-t\vb\cdot\vnabla\psi
      =(i\vb\cdot\vx)\psi-(\theta/2)\vb\times\vnabla\psi
      -t\vb\cdot\vnabla\psi
      \label{fimp}
 \end{equation}
 and
  \begin{equation}
 \psi_{\vb}^{\star}\left(\vx, t \right) = e^{- \imath \left(
 \frac{\vb^2}{2}t  \right)}e^{\imath
 \vb\cdot \vx}_{\star} \star \psi\left(\vx-\vb t, t\right), \label{MoyTr}
 \end{equation}
 respectively, where $ e^{\imath
 \vb\cdot \vx}_{\star}$ is the exponential w.r.t. the $\star$ product.
 Analogously, a $\theta$ - deformed expansion symmetry   is allowed.
 In infinitesimal form, it is given as \cite{HMS1}
 \begin{eqnarray}
 \delta^*_{\eta}\vx=\eta t\vx,\; \delta^*_{\eta}t=\eta t^2,\;
 \delta^*_{\eta}\psi=-\eta\big[(-\frac{i}{2}x^2+t)\psi
 +t\vx\cdot\vnabla\psi+t^2\p_{t}\psi\big] -\eta\left[
 \frac{\theta}{2}\vx\times\vnabla\psi+\frac{\theta^2}{4}
 \p_{t}\psi\right], \label{exexp}
 \end{eqnarray}
 and the usual dilations \cite{JHN}
 \begin{equation}
 \delta_{\Delta}\vx=\Delta\vx,\qquad
     \delta_{\Delta}t=2\Delta t,\qquad
     \delta_{\Delta}\psi=
     -\Delta\big[\psi+\vx\cdot\vnabla\psi+2t\p_{t}\psi\big].
     \end{equation}
 where $\eta$ and $\Delta>0$ are real parameters.

 The Noether theorem still holds and the associated
 conserved quantities may get new terms,
 \begin{equation}
     M= \!\int\! d^2x\vert\psi\vert^2 ,\,{\emph{H}}_{0}=\int d^2\vx\,
      \frac{1}{2}\vert\vnabla\psi\vert^2,
      \;{{\emph{P}}}_{i}=-i\displaystyle\int
 d^2x\bar{\psi}\p_{i}\psi, \;
      {\emph{J}}=-i\int d^2\vx\epsilon_{ij}x_{i}\bar{\psi}\p_{j}\psi,
 \label{enerangmommass}
 \end{equation}
 \begin{equation}
      {\emph{G}}_{i}=-\int d^2\vx\,x_{i}\vert\psi\vert^2
      +t{{\emph{P}}}_{i}
      +\frac{\theta}{2}\epsilon_{ij}\,{\emph{P}}_{j}
      \label{exoboost}
 \end{equation}
 \begin{equation}
      \begin{array}{ll}
      {\emph{D}}=-2t{\emph{H}}_{0}+\displaystyle\frac{1}{2i}\int d^2\vx x_{i}
      \big(\bar{\psi}\p_{i}\psi-(\p_{i}\bar{\psi})\psi\big),\;
      {\emph{K}}=
      t^2{\emph{H}}_{0}+t{\emph{D}}-
      \displaystyle\frac{1}{2}\int\!d^2\vx\,\vx^2|\psi|^2
      +\frac{\theta}{2}{\emph{J}}-\frac{\theta^2}{4}\emph{H}_{0}.
      \label{ncconf}
      \end{array}
 \end{equation}
 Since  we can use the usual Poisson brackets
 $\big\{\psi(\vx,t),\bar{\psi}(\vx',t')\big\} =
 -i\delta(\vx-\vx')$, the symmetry algebra, expressed in terms of
 the above generators, contains additional $\theta$ - dependent
 terms, as in
  (\ref{exoticcharge}). One establishes  the relation
   \begin{equation}k= \theta \emph{M}. \label{ExoCh}
 \end{equation}
  The other modified  brackets are
  \begin{equation}
      \{{\emph{K}},{\emph{G}}_{i}\}=\theta\epsilon_{ij}{\emph{G}}_{i},
      \;
      \{{\emph{D}},{\emph{K}}\}=-2{\emph{K}}+\theta{\emph{J}}-\theta^2{\emph{H}}
 _{0},\;
      \{{\emph{D}},{\emph{G}}_{i}\}=-{\emph{G}}_{i}+\theta\epsilon_{ij}{\emph
 {P}}_{j}.
      \label{mixed}
 \end{equation}
 All other commutation relations among the symmetry generators
 remain the same as in the commutative case \cite{JHN}.

 Turning  to the interacting theory, one can observe that
 all the self-interactions can be written in terms of the "chiral"
 densities $\rho_{+}=\bar{\psi}\star\psi$, and
 $\rho_{-}=\psi\star\bar{\psi}$.

 Now, if the usual implementation of the Galilei is applied, the
 densities  transform infinitesimally as
 $\delta^{0}_{b}\rho_{\pm}=\pm
 \frac{\theta}{2}\vb\times\vnabla\rho_{\pm}
 -t\vb\cdot\vnabla\rho_{\pm}$. Analogously, resorting to the
 "exotic" boost (\ref{MoyTr}), they change according to
 $$
 \delta^{\star}_{b}\rho_{+}=-t\vb\cdot\vnabla\rho_{+}, \;
 \delta^{\star}_{b}\rho_{-}= -t\vb\cdot\vnabla\rho_{-}
 -\theta\vb\times\vnabla\rho_{-}.
 $$  
 Consequently, the variation of
 a generic potential  $V$ becomes exact w.r.t. such a
 transformation if it only depends
  on one type of $\rho_{\pm}$:
 \begin{equation}
      \delta^*_{b}\widetilde{V}_{+} = -t\vb\cdot\vnabla\widetilde{V}_{+},\;
 \delta^*_{b}{\widetilde{V}_{-}} = -t\vb\cdot\vnabla
 \widetilde{V}_{-} -\theta\vb\times\vnabla\widetilde{V}_{-}.
 \end{equation}
      In conclusion, any  ``pure'' expression
 $ V_{\pm} = V(\rho_{\pm}) $ provides us with a theory which is {\it
 Galilei-invariant} both in the the conventional and
  the ``$\star$-implementation''. Chiral potentials of this kind
 were considered by several authors \cite{Langmann}.

 Finally, one can verify
 by direct computation that any potential made of products of
 $\rho_{\pm}$  necessarily breaks the conformal invariance. This is
 a consequence of the non local character of the theory and it
 could be related to the so - called UV/IR mixing \cite{Szabo}.

 Analyzing now the symmetry properties of the  gauged model
 (\ref{Clag}) with the pure fourth order self-interaction $ V =
 \lambda \lf \psi \star \bar{\psi} \rg^2$ \cite{LMS}, we see that
 the equations of motion (in Moyal and operatorial form, respectively)
 \begin{equation}
   \begin{array}{lll}
      \imath D_{t}\psi+\frac{1}{2}{\vD}^2\psi + \lf 2 \lambda-
      \displaystyle\frac{1}{2\kappa}\rg  \psi \star \bar{\psi}\star\psi
      =0,
        &  \quad & \imath D_{t}\hp+\frac{1}{\theta} D \Db \hp +
       \lf 2 \lambda  -\frac{1}{2\kappa}\rg \hp \hpd \hp =  0,
       \label{NCNLS}
       \\[8pt]
     \kappa E_{i}-\varepsilon_{ik}j_{-, k} = 0 , & \textrm{or}\quad &
     2 \kappa \imath D_t K  =  \left\{K, \hp \hpd \right\} - 2\hp \hc
     \hpd,
 \label{NCFCI}
 \\[8pt]
 \kappa B+ \rho_{-} = 0, &  \quad & \lq K , K^\dagger \rq  =  1
 -\displaystyle\frac{\theta}{\kappa} \hp \hpd,
 \label{NCGauss}
   \end{array}
 \end{equation}
 with $B=\epsilon_{ij}F_{ij} $, $E_{i}=F_{i0}$ and
 ${\vj}\strut{\,}_-=\frac{1}{2i}\left(\vD\psi\star\bar\psi
      -\psi\star(\overline{\vD\psi})\right)$,  possess an
 explicit ``chirality'' in the Hall  and the Gauss law, respectively.
 Both the usual
 and the $\delta^*$ (\ref{MoyTr} -\ref{fimp})  implementations
 break the invariance of the Gauss law  under a Galilean boost.
  However, Galilean
 symmetry is restored if we consider  the {\it
 antifundamental representation}
 \begin{equation}
      \delta_{*}\psi=\psi\star(i\vb\cdot\vx)-t\vb\cdot\vnabla\psi
      =
      (i\vb\cdot\vx)\psi+\frac{\theta}{2}\vb\times\vnabla\psi
      -t\vb\cdot\vnabla\psi,
      \label{afimp1}
 \end{equation}
 which is obtained as a right action of the $U(1)_*$ group, or more
 simply, changing  the sign of $\theta$ in (\ref{fimp}). At the
 same time, the gauge field $A_{\mu}$ transforms as usual by \beq
 \delta_{}A_{i}=-t\vb\cdot\vnabla A_{i}, \;
      \delta_{}A_{0}=-\vb\cdot\vA-t\vb\cdot\vnabla A_{0}. \label{afimp2} \eeq

      Finally, it is  straightforward to prove that "pure" chiral
      self-interactions
 $V_{\pm}$ are also invariant under the $\delta_*$ transformations.
 Thus, the required galileian symmetry strongly selects the type of
 interaction, as we considered in (\ref{NCNLS}).

 On the other hand, any potential breaks the conformal
 invariance.

      A remarkable consequence is that one can build up the new conserved
 Noether
      charges associated to the boost charge
 \begin{equation}
 \vG^r=t\vP-\int\!\vx\rho_+\, d^2\vx . \label{Gr}\end{equation}
 Their Poisson brackets differ from (\ref{exoticcharge}) only for
      the sign of $\theta$.

 One could actually define the family of conserved quantities
 $G_{i}^{\lf\alpha\rg}=G_{i}^r+\frac{\alpha}{2}\epsilon_{ij}P_{j}$,
 parametrized by a real $\alpha$. This leads to  new transformation
 rules $
 \delta^{\lf\alpha\rg}\psi=\vb\cdot\big\{\psi,\vG^{\alpha}\big\}$
 and $\delta^{\lf\alpha\rg} A_i =\vb\cdot\big\{A_i
 ,\vG^{\alpha}\big\}$, depending on $\alpha$. The Poisson brackets
 for the boost charges are
 \begin{equation} \big\{G_{i}^{\alpha}, G_{j}^{\alpha}\big\}=
      \epsilon_{ij}(\alpha-\theta)\!\int\!\vert\psi\vert^2 d^2x
      \label{rncboost}
      \end{equation}
     So that, for $\alpha=0$ we recover the $^*$-implementation
 (\ref{exoticcharge}). For $\alpha=\theta$ instead, we act on the
 matter field as in the commutative case, because of  vanishing of  the
 second central charge.  But the gauge potential changes non-conventionally.
 However, following \cite{SW},
   the gauge fields in the non commutative
 ($\theta\neq0$) domain must be related to the commutative
 ($\theta=0$) case by a differential relation in $\theta$,
 precisely by
 \begin{equation}
 \frac{\ \p}{\p\theta}A_{i}(\theta)=-\frac{1}{4}\epsilon_{k l}
 \Big(A_{k}\star (\p_{l}A_{i}+F_{li})+(\p_{l}A_{i}+F_{li})\star
 A_{k}\Big). \label{SWcond}
 \end{equation}
 This equation is manifestly form-invariant w. r. t.  boosts,
  provided $\alpha$ does not depend on $\theta$. On
 the other hand, the boost transformation for the gauge fields on
 the ordinary plane ($\theta = 0$) holds only for $\alpha = 0$. In
 conclusion, the boost generator (\ref{Gr}) is the only admissible one,
 because it is continuous for $\theta \rightarrow  0$. Hence, the
 non trivial second charge is dynamically defined by
 (\ref{rncboost}) for $\alpha = 0$.

 However,
 the proposed infinitesimal
 transformations (\ref{afimp1} - \ref{afimp2}) are not gauge covariant. This
 fact may provide
 difficulties in deriving correct gauge invariant conserved
 quantities and invariant configurations under specific symmetry
 subgroups, as it is well known in dealing with gauge field
 theories. But, resorting to this analogy, in correspondence to an
 infinitesimal linear coordinates transformation $\delta_f x^\mu =
 -f^\mu$ of the NC - plane, in \cite{JPct} it was proposed to
 associate an infinitesimal gauge-covariant transformation on a
 vectorfield  $\hat{A}_\mu $ by $ \hat{\delta}_f \hat{A}_\mu =
 \half \lGr \hat{f}^\nu, \hat{F}_{\nu \mu}\rGr_*$, where the Moyal
 - anti-commutator $\lGr \cdot, \cdot\rGr_*$ has been introduced.
  To be specific, we modify the expressions
 (\ref{afimp1})-(\ref{afimp2}) in the gauge - covariant way by \beq
      \hat{\delta}_\vb\psi=\psi\star(\imath\vb\cdot\vx)-t\vb\cdot\vD\psi,
       \quad \hat{\delta}_\vb A_{i} = -t b_j F_{j i} = t \varepsilon_{i j}b_j
 B, \;
      \hat{\delta}_\vb A_{0}=-t\vb \cdot \vec{E}.
     \label{covboost} \eeq
 Of course, these transformations can be expressed in terms of
 $\rho_-$ and $\vj_-$, accordingly to (\ref{NCNLS}). It is
 remarkable that their algebra closes up to a gauge
 transformations. Specifically, noticing that
 \begin{equation}
 \hat{\delta}_\vb B = -t \vb \cdot \vD B, \; \hat{\delta}_\vb E_i =
 - t \vb \cdot \vec{D} E_i - \varepsilon_{i j}b_j B,
 \label{CovTrasFieldStr}
 \end{equation}
 one can find
 \begin{eqnarray}
  \lq
 \hat{\delta}_\vb,\hat{\delta}_{\vb'}\rq \psi &=& \imath \vb \times
 \vb' \lf  \theta - t^2 B \rg \psi  ,\\[6pt]
 \nonumber \lq
 \hat{\delta}_\vb, \hat{\delta}_{\vb'}\rq A_i &=&-t^2 \vb \times
 \vb'  D_i B,
 \\[6pt]
 \nonumber
 \lq \hat{\delta}_\vb, \hat{\delta}_{\vb'}\rq A_0 &=&
 -\vb \times \vb'\lf t^2 D_0 B +  2 t B \rg.
 \end{eqnarray}
 Thus, the algebra of the covariant boosts closes up to a gauge transformation
 generated by
 $  -\vb \times \vb' \;t^2  B$. Moreover, the matter field
 acquires a time independent  phase factor  plus the gauge transformation
 contribution.
   Because of this result,  the
 integration of such infinitesimal transformations is
 in general prevented.

 However, the integrability can be  obtained in the special case
 $D_{\mu} B =0$. Because of the Gauss law, also the covariant time
 derivative of the chiral density $\rho_-$ is vanishing, Thus,
 since a covariant version of the continuity equation holds, the
 chiral current $\vj _-$ is covariantly solenoidal, i.e. $\vD \cdot
 \vj_- =0$. Finally, taking into account the field - current
 identity, we end up with the covariant irrotational condition for
 the electric field $\vD \times \vE =0$, consistently with the
 Bianchi identity for the connection $A_{\mu}$.

 Although the above conditions restrict us to special field
 configurations,  they still allow for non trivial solutions.

 Let us start with static solutions of (\ref{NCFCI}) in the BPS
 limit $2 \lambda \kappa = 1$ \cite{LMS}. They are given by

 \begin{equation}
 \hp = \sqrt{\frac{\kappa}{\theta}} | 0 \rangle \langle \zeta_0 | ,
 \qquad K =  \zeta_0  | 0 \rangle \langle 0 | + S_1 \hc
 S_1^{\dagger}, \qquad A_0 = - \frac{1}{2 \kappa}| 0 \rangle
 \langle 0 |, \label{HSvorstat} \end{equation}
 where $S_1 =\sum_{i
 = 0}^{\infty}| i \rangle \langle i+1|$.
 Here $ | 0 \rangle $ is the Fock vacuum $\hc  | 0 \rangle  = 0$ and $ |\zeta_ 0
 \rangle $ is a coherent state centered at the position $\zeta_0$, i.e.
 \beq
  |\zeta_ 0 \rangle = e^{- \half |\zeta_0|^2} e ^{\zeta_0 \hcd}    | 0 \rangle.
 \eeq
 The Gauss law give us  $B = - \frac{1}{\theta}| 0
 \rangle \langle 0 |$, i.e. we have a field configuration from the class $D_
 {\mu} B =0$ and
 $\vE = 0$.

 Now, time dependent solutions are obtained by acting on the static solutions
 (\ref{HSvorstat}) by a finite gauge covariant boost (with velocity $v$).
 In fact, substituting the static solution into the infinitesimal
 transformations
 (\ref{covboost}) and (\ref{CovTrasFieldStr}), one obtains
 \beq
 \hp = \sqrt{\frac{\kappa}{\theta}} | 0 \rangle \langle \phi\lf t \rg | \label
 {HSmov},
 \eeq
 where
 \beq
 | \phi\lf t \rg  \rangle = e^{\lf t - \imath \theta \rg \hcd - \bar{v} \lf t +
 \imath \theta \rg +
 t \lf \bar{v} \zeta_0 - v \bar{\zeta_0} \rg} |\zeta_0 \rangle.
 \eeq
 Manipulations of this expression,  resorting to the Baker - Campbell -
 Hausdorff formula, lead to
 \beq
 | \phi\lf t \rg  \rangle = e^{\imath \alpha\lf t \rg} |\zeta_0 + v \lf t -
 \imath \theta \rg \rangle,
 \eeq
 which represents a coherent state,
 uniformly moving, with velocity $v$. The time dependent phase
 factor $\alpha\lf t \rg$ has the expression
 \beq
 \imath \alpha \lf t \rg  = -\imath \frac{\theta}{2}  \lf \bar{v} \zeta_0 +  v
 \bar{\zeta_0} \rg +
 \frac{t}{2} \lf \bar{v} \zeta_0 - v \bar{\zeta_0} \rg.
 \eeq
 For the gauge component $K$ one obtains the expressions
 \beq
 K \lf t \rg) = \lf \zeta_0 + v t \rg  | 0 \rangle \langle 0 | + S_1 \hc S_1^
 {\dagger},
 \eeq
 while $A_0$ remains static. Note that $B$ is left invariant, i.e. it still
 belongs to the class $D_{\mu}B = 0$.
  Solutions of the kind (\ref{HSmov}) were found in
 \cite{HS}. From (\ref{CovTrasFieldStr}) we se that
  the CS-electric field takes the value $
 E_i = \frac{1}{\theta}\varepsilon_{i j} v_j |0 \rangle
 \langle 0 |$. Notice that this field configuration is similar as in
 the Hall motion. In particular, the relation $\frac{|E|}{|B|} = v$
  holds, while the particle is moving
   in the direction orthogonal to $\vec{E}$.

 Finally, in the BPS limit $2 \lambda \kappa = 1$, one obtains a simple
 expression for
  energy of the particle interacting  with the Chern - Simons
 field and with itself. In fact, one has
 $ \mathcal{E} = - 2 \pi {\rm Tr} \lq \hpd \Db D \hp \rq$. For the
 solution (\ref{HSmov}), one sees that $\mathcal{E} = 2 \pi \kappa
 \theta |v|^2 = \half \theta M  |v|^2$, which
 confirms the continuous spectrum of the vortex energy. More
 interesting it is its expression as kinetic energy for a particle
 possessing the second central charge (\ref{ExoCh}) as mass and
 moving at the selected Hall velocity, found by the ratio
 $|E|/|B|$.

  In conclusion, we have discussed the symmetries of a field theory
  on the non commutative plane. The first result is that we can
  recover galilean invariant theories, but only  "chiral"
  interactions are admissible. However, the boost generators no longer commute,
  but they close up by the introduction of the second central
  extension parameter $\kappa$, which is dynamically determined by
  the particle mass and the non commutativity parameter of the
  plane. Finally, classes of uniformly moving solutions can be
  built by resorting to freely moving coherent states.
  However, the associated CS-magnetic and electric field are not
  trivial at all, even if they result to be static. More general
  type of moving solutions can be found \cite{Had} and then boosted. In
  particular, one can show that two vortex solutions can be boosted,
  without changing their relative positions.
  This is in contrast with the
  dipolar solutions found by \cite{Baketal}.

  This work was partially supported
  by the Murst (grant SINTESI 2002) and by the INFN (Iniz. Spec.
  Le41). Two of us (PAH and  LM) acknowledge the
  organizers for their hospitality at the  International Workshop
 {\it Nonlinear Physics}. Gallipoli'2004.



\begin{thebibliography}{99}

 \bibitem{Szabo}M. R. Douglas, N.A. Nekrasov, {\it Rev. Mod. Phys}
 {\bf 73} (2001) 977.  R.~J. Szabo, {\it Quantum Field Theory on
 noncommutative spaces}. {\it Phys. Rep.} {\bf 378}(2003),  203,
 [\texttt{hep-th/0109162}] and refereces therein.
 \bibitem{Laughlin}
 R.B. Laughlin, {\it Phys. Rev. Lett.} {\bf 50} (1983), 1395.
 \bibitem{Arovas}
 D. Arovas, J.R. Schrieffer, F. Wilczek, {\it Phys. Rev. Lett.} {\bf
 53} (1984), 772.
 \bibitem{Girvin}
 S.M. Girvin, T. Jach, {\it Phys. Rev. } {\bf A 29} (1984), 5617.
 \bibitem{Niu}
  M.-C. Chang and Q. Niu,  {\it Phys. Rev. } {\bf B 53} (1996),
 7010.
 \bibitem{HMSanomal}
 P.A. Horv{\'a}thy, L. Martina, P.C. Stichel,"{\it Enlarged Galilean
 symmetry of anyons and the Hall Effect}". \texttt{hep-th/0412090}
 \bibitem{Bohm}
 A. Bohm, A. Mostafazadeh, H. Koizumi, Q. Niu, J. Zwanziger: {\it
 The Geometric Phase in Quantum Systems}, Springer - Verlag
 (Berlin, 2003), and refereces therein.
 \bibitem{LSZ}
 J.~Lukierski, P.~C.~Stichel, W.~J.~Zakrzewski,
    {\sl Annals of Physics (N. Y.)} {\bf 260}, 224 (1997), and
    {\sl ibid}. {\bf 306}, 78 (2003) [\texttt{hep-th/0207149}].
 \bibitem{DH}
 C.~ Duval and P.~A.~Horv{\'a}thy, {\sl Phys. Lett.} {\bf B 479},
 284 (2000) [\texttt{hep-th/0002233}]; {\sl J. Phys.} {\bf A 34},
 10097 (2001) [\texttt{hep-th/0106089}];
 P.~A.~Horv{\'a}thy, {\sl Ann. Phys.} (N. Y.) {\bf 299},
 128-140 (2002). [\texttt{hep-th/0201007}].
 \bibitem{exotic}
 J.-M.~L\'evy-Leblond, {\it Galilei group and Galilean
 invariance}. in {\it Group Theory and Applications} (Loebl Ed.),
 {\bf II}, Acad. Press, New York, p. 222 (1972); A. Ballesteros,
 N.~Gadella and M.~del Olmo, {\sl Journ. Math. Phys.} {\bf 33},
 3379 (1992); Y.~Brihaye, C.~Gonera, S.~Giller and P.~Kosi\'nski,
 \texttt{hep-th/9503046} (unpublished); D.~R.~Grigore, {\sl Journ.
 Math. Phys.} {\bf 37}, 240 and
    {\sl ibid}. {\bf 37} 460 (1996).
 \bibitem{Susskind}
 L. Susskind, hep-th/0101029 (2001)
 \bibitem{LMS}
 G.~S. Lozano, E. F. Moreno, F. A. Schaposnik, {\sl Journ. High
 Energy Phys}. {\bf 02}  036 (2001) [\texttt{hep-th/0012266}]. See
 also the review by F. A. Schaposnik, \texttt{hep-th/0408132}.
 \bibitem{BakCS}
 D.~Bak, S.~K.~Kim, K.-S. Soh, and J. H. Yee, {\it Noncommutative
 Chern-Simons solitons}. { Phys. Rev}. {\bf D64}, 025018 (2001)
 \bibitem{kappaquant}
 D.~Bak, K.~Lee, J.-H. Park, {\it Chern-Simons theories on the
 noncommutative plane}. { Phys. Rev. Lett}. {\bf 87}, 030402
 \bibitem{Baketal}
 D.~Bak, S.~K.~Kim, K.-S. Soh, and J. H. Yee, { Phys. Rev. Lett}.
 {\bf 85}, 3087 (2000).
 \bibitem{HMS1}
 P.A. Horv{\'a}thy, L. Martina, P.C. Stichel, {\it Phys. Lett.} {\bf
 B 564} (2003), 149.
 \bibitem{HMS2}
 P.A. Horv{\'a}thy, L. Martina, P.C. Stichel, {\it Nucl. Phys.} {\bf
 B 673} (2003), 301.
 \bibitem{JHN}
 R.~Jackiw, {\sl Physics Today} {\bf 25}, 23 (1980); U.~Niederer,
 {\sl Helvetica Physica Acta} {\bf 45}, 802 (1972); C.~R.~Hagen,
 {\sl Phys. Rev}. {\bf D5}, 377 (1972).
 \bibitem{HS}
 P.A. Horv{\'a}thy,  P.C. Stichel, {\it Phys.Lett.} {\bf B 583}
 (2004), 353.
 \bibitem{Had}
 L. Hadasz, U. Lindstrom, M. Rocek, R. von Unge, {\it Phys.Rev.}
 {\bf D 69 }(2004), 105020.
 \bibitem{Langmann}
   E.~Langmann, R.~J.~Szabo,
 {\sl Phys. Lett.} {\bf B533}, 168 (2002);
   E.~Langmann, {\it Nucl. Phys}. {\bf B654}, 404 (2003);
 E.~Langmann, R.~J.~Szabo, and K. Zarembo, \texttt{hep-th/0308082}.
 \bibitem{SW}
 N. Seiberg and E. Witten, {\it J. High Energy Phys.} {\bf
 09}(1999), 032.
 \bibitem{JPct} R. Jackiw, S. -Y. Pi, {\it Phys. Rev.
 Lett.} {\bf 88} (2002), 111603-1.
 \end{thebibliography}
 \end{document}